\documentclass[a4]{article}
\usepackage{graphics}
\begin{document}

\title{Evolution equations for truncated moments of the parton distributions}
\author{Dorota Kotlorz\footnote{Opole University of Technology,
 Division of Physics, Ozimska 75, 45-370 Opole, Poland, e-mail:
{\tt dstrozik@po.opole.pl}} \and Andrzej Kotlorz\footnote{Opole University
of Technology, Division of Mathematics, Luboszycka 3, 
45-036 Opole, Poland, e-mail: {\tt kotlorz@po.opole.pl}}}
\date{October 13, 2006}
\maketitle

\abstract{
We derive evolution equations for the truncated Mellin moments of the parton
distributions.
We find that the equations have the same form as those for the partons
themselves. The modified splitting function for n-th moment
$P'(n,x)$ is $x^{n}P(x)$, where $P(x)$ is the well-known splitting function
from the DGLAP equation.
The obtained equations are exact for each n-th moment and for every truncation
point $x_0\in (0;1)$. They can be solved with use of standard methods of
solving the DGLAP equations.

This approach allows us to avoid the problem of dealing with the unphysical
region $x\rightarrow 0$. Furthermore, it refers directly to the physical
values - moments (rather than to the parton distributions), what enables one to
use a wide range of deep-inelastic scattering data in terms of smaller number
of parameters.

We give an example of an application.\\ \\
PACS {12.38.Bx} {Perturbative calculations}, {11.55.Hx} {Sum rules}, 
{Truncated moments}

\section{Introduction}
\label{intro}
QCD interactions between partons violate the Bjorken scaling \cite{b1}. The
quark and the gluon distribution functions change with $Q^2$ according to
the well-known Dokshitzer-Gribov-Lipatov-Altarelli-Parisi (DGLAP) evolution
equations \cite{b2}-\cite{b5}. The DGLAP equations can be solved with use of
either the Mellin transform or the polynomial expansion in the $x$-space.
The differentio-integral Volterra-like evolution equations change after
the Mellin transform into simple differential and diagonal ones in the moment
space and can be solved analytically. Then one can again obtain the $x$-space
solutions via the inverse Mellin transform. The only problem is knowledge of
the input parametrisation for the whole region $0\leq x \leq 1$ what is
necessary in the determination of the initial moments of the distribution
functions. Using the polynomial expansion method, one deals with the parton
distributions only in a limited range of the Bjorken variable:
$x\leq z\leq 1$. This is very important because of the experimental
constraints. The lowest value of $x$ in present experiments is about $10^{-5}$
and the limit $x\rightarrow 0$, which implies that the invariant
energy $W^2$ of the inelastic lepton-hadron scattering becomes infinite
($W^2=Q^2(1/x-1)$), will never be attained. Thus the polynomial expansion
technique allows one to avoid the problem of the unphysical region.

An important role in DIS analyses play different sum rules, which refer to the
moments of the parton distributions. One can compare experimental estimations
for the moments of structure functions with theoretical predictions. Because
experimental data do not cover the whole region of $x$, it is very useful to
consider in the theoretical approach truncated moments of structure functions.
The DGLAP evolution equations for the truncated moments have been discussed in
\cite{b6}-\cite{b10}. The presented methods, based on the approximate formulae
are valid only for very small value of the truncation point ($x_0\leq 10^{-2}$)
and suffer from large errors as $x_0\geq 0.1$. Besides, the not diagonal
evolution equations for truncated moments, obtained in \cite{b6}-\cite{b8},
can be solved with a satisfactory precision for $n\geq 2$, while in a case of
the first moment the precision is significantly worse.

In this paper we derive the evolution equations for the truncated moments of
the parton distributions, which are exact for every value of $x_0$ and for
each $n$-th moment. We find that the truncated moments satisfy the DGLAP
evolution with a modified splitting function $P(x)\rightarrow x^n P(x)$ in
the Mellin convolution. This approach enables one to avoid uncertainties from
the unmeasurable $x\rightarrow 0$ region and also to use the smaller number of
the input parameters.

The content of this paper is as follows. In Sect.2 we present the DGLAP
evolution equations for the truncated moments of the parton distributions.
A detailed derivation is given in the appendix. An example of an application to the
determination of the contribution to the Bjorken sum rule is contained in
Sect.3. Finally, Sect.5 contains a summary and conclusions.

\section{Derivation of the evolution equations for the truncated Mellin moments
of the parton densities}
\label{sec:2}

The $Q^2$ evolution of the quark and the gluon densities is described by
the well-known DGLAP equations \cite{b2}-\cite{b5}. Each equation of this
system has the form
\begin{equation}\label{eq.2.1}
\frac{dq(x,t)}{dt}=\frac{\alpha_s(t)}{2\pi}\; (P\otimes q)(x,t).
\end{equation}
In the above formula $q(x,t)$ denotes the parton distribution,
\begin{equation}\label{eq.2.2}
t= \ln\frac{Q^2}{\Lambda_{QCD}^2},
\end{equation}
$\alpha_s(t)$ is the running coupling and $\otimes$ denotes the Mellin
convolution:
\begin{equation}\label{eq.2.3}
(A\otimes B)(x)\equiv\int\limits_{x}^{1} \frac{dz}{z}\,
A\left(\frac{x}{z}\right)\,B(z).
\end{equation}
The splitting function $P(z,t)$ can be expanded in the perturbative series of
the $\alpha_s(t)$.\\
The $n$-th Mellin moment of the function f(x) is defined as
\begin{equation}\label{eq.2.4}
\bar{f}_{n}=\int\limits_0^1 dx\, x^{n-1}\, f(x).
\end{equation}
Taking into account the relation
\begin{equation}\label{eq.2.5}
\int\limits_0^1 dx\, x^{n-1}\,(A\otimes B)(x)=
\bar{A}_n\,\bar{B}_n,
\end{equation}
one can find that the moments of the parton distributions obeys the evolution
equation
\begin{equation}\label{eq.2.6}
\frac{d\bar{q}_{n}(t)}{dt}=\frac{\alpha_s(t)}{2\pi}\;
\gamma_n(t)\, \bar{q}_{n}(t),
\end{equation}
where the anomalous dimension $\gamma_n(t)$ is a moment of the splitting
function P(z,t)
\begin{equation}\label{eq.2.7}
\gamma_n(t)=\int\limits_0^1 dz\, z^{n-1}\, P(z,t).
\end{equation}
Eq.(\ref{eq.2.6}) can be solved analytically and the parton density $q(x,t)$
can be found via the inverse Mellin transform
\begin{equation}\label{eq.2.8}
q(x,t)=\frac{1}{2\pi i}\int\limits_{c-i\infty}^{c+i\infty} dn\,
x^{-n}\, \bar{q}_n(t).
\end{equation}
The only problem is the knowledge of the input parametrisation for the whole
region $0\leq x\leq 1$, what is necessary in the determination of the initial
moments $\bar{q}_n(t=t_0)$.

Using the truncated moments approach one can avoid the uncertainties from the
region $x\rightarrow 0$, which will never be attained experimentally. If we
truncate the Mellin moment at $x_0$, we obtain
\begin{equation}\label{eq.2.9}
\bar{f}_{n}(x_0)=\int\limits_{x_0}^1 dx\, x^{n-1}\, f(x).
\end{equation}
The evolution equation (\ref{eq.2.1}) implies for the truncated moment of $q$
the following formula:
\begin{equation}\label{eq.2.10}
\frac{d\bar{q}_n(x_0,t)}{dt}=
\frac{\alpha_s(t)}{2\pi}\int\limits_{x_0}^1 dx\, x^{n-1}\,(P\otimes q)(x,t),
\end{equation}
where
\begin{equation}\label{eq.2.10a}
\bar{q}_{n}(x_0,t)=\int\limits_{x_0}^1 dx\, x^{n-1}\, q(x,t).
\end{equation}
We derive in the appendix an interesting relation for the truncated moments,
namely
\begin{equation}\label{eq.2.11}
\int\limits_{x_0}^1 dx\, x^{n-1}\,(P\otimes q)(x)= (P'\otimes \bar{q}_n)(x_0),
\end{equation}
with
\begin{equation}\label{eq.2.12}
P'(n,z)= z^n\, P(z)
\end{equation}
Hence eq.(\ref{eq.2.10}) can be written as
\begin{equation}\label{eq.2.13}
\frac{d\bar{q}_n(x_0,t)}{dt}=
\frac{\alpha_s(t)}{2\pi}\; (P'\otimes \bar{q}_n)(x_0,t).
\end{equation}
In this way we have obtained the DGLAP evolution equation for the truncated
moments $\bar{q}_n(x_0,t)$, very similar to the original equation
(\ref{eq.2.1}) for the partons themselves. In our approach, $P'(n,z)$
from (\ref{eq.2.12}) plays a role of the splitting function for truncated
moments.

Now we can adapt to eq.(\ref{eq.2.13}) known methods of solving the DGLAP
evolution equation e.g. \cite{b11},\cite{b12}.
Dealing with the truncated moments allows us to study their evolution without
making any assumption on the small-$x$ behaviour of the parton densities. One
needs to know only the truncated moments of the parton distributions at the
initial scale $t_0$ (e.g. from the experimental data), what constrains a number
of the input parameters. In our approach we can also test different parton
parametrisations comparing the theoretical predictions for sum rules, which
involve $\bar{q}_n(x_0,t)$, with the experimental data.

\section{An example of an application}
\label{sec:3}

Eq. (\ref{eq.2.13}) can be rewritten in the full form
\begin{equation}\label{eq.2.14}
\frac{d\bar{q}_n(x_0,t)}{dt}=\frac{\alpha_s(t)}{2\pi}
\int\limits_{x_0}^{1}\frac{dz}{z}\,P'\left(n,\frac{x_0}{z},t\right)\,
\bar{q}_n(z,t),
\end{equation}
where
\begin{equation}\label{eq.2.15}
P'\left(n,\frac{x_0}{z},t\right)=
\left(\frac{x_0}{z}\right)^n P\left(\frac{x_0}{z},t\right).
\end{equation}
In order to solve eq.(\ref{eq.2.14}), we use the Chebyshev polynomial
expansion of $P'$ and $\bar{q}_n$.
The Chebyshev polynomials technique \cite{b13} was successfully used by 
J.Kwieci\'nski in many QCD treatments e.g. \cite{b14}-\cite{b17}. Using this
method one obtains the system of linear differential equations instead of the 
original integro-differential ones. The Chebyshev expansion provides a robust 
method of discretising a continuous problem. More detailed description of the
Chebyshev polynomials technique is given e.g. in the appendix of \cite{b10}.
The Chebyshev polynomial approach in the case of the truncated moments is
more effective than in the case of the parton densities themselves.
This is because $n$-th moments for $n\geq 1$ are usually nonsingular, when
$x_0\rightarrow 0$.
Thus we can obtain reliable results for every value of $n\geq 1$, $x_0$ and
also $Q^2$, independently of the input parton distributions.

Here we would like to present solutions of eq.(\ref{eq.2.13}) for the
nonsinglet polarised structure function $g_1$. In this way we are able to
determine the contribution to the Bjorken sum rule \cite{b18},\cite{b19}.
For simplicity we use LO approximation. In our case
\begin{equation}\label{eq.2.16}
q(x,t)=g_1^{NS}(x,t)=g_1^p(x,t)-g_1^n(x,t),
\end{equation}
so
\begin{equation}\label{eq.2.17}
\bar{q}_{n}(x_0,t)=\int\limits_{x_0}^1 dx\, x^{n-1}\, g_1^{NS}(x,t).
\end{equation}
The results for $\bar{q}_{n}(x_0,t)$ (\ref{eq.2.17}) are shown in Figs.1-3.
In Fig.1 we plot the truncated moments as a function of $x_0$.
We use two different inputs of $g_1^{NS}(x,t_0)$ at the low scale
$Q_0^2=1\; {\rm GeV}^2$:
\begin{eqnarray}\label{eq.2.18}
g_1^{NS}(x,t_0) &\sim& (1-x)^3, \\ \label{eq.2.19}
g_1^{NS}(x,t_0) &\sim& x^{-0.5}(1-x)^{3}.
\end{eqnarray}
The evolution scale $Q^2=10\; {\rm GeV}^2$. 
Figs.2, 3 show the $Q^2$ dependence of the (\ref{eq.2.17}) at $x_0=0.01$ and
$x_0=0.1$, respectively. Here we consider again the both parametrisations
(\ref{eq.2.18}),(\ref{eq.2.19}).
We can also determine the contribution to the Bjorken sum rule
\begin{equation}\label{eq.2.20}
\Delta I_{BSR}(x_1,x_2,t)=
\int\limits_{x_1}^{x_2} dx\, g_1^{NS}(x,t).
\end{equation}
For $x_1=0.003$, $x_2=0.7$ and $Q^2=10\; {\rm GeV}^2$ we obtain
$6\Delta I_{BSR}=1.23$ and 1.07 for the inputs (\ref{eq.2.18}) and
(\ref{eq.2.19}), respectively. These results can be compared to the SMC data
\cite{b20}, which yield $6\Delta I_{BSR}(0.003,0.7,10)=1.20\pm 0.24\pm 0.15$.
\begin{figure}
\begin{center}
\resizebox{0.7\textwidth}{!}{
\includegraphics{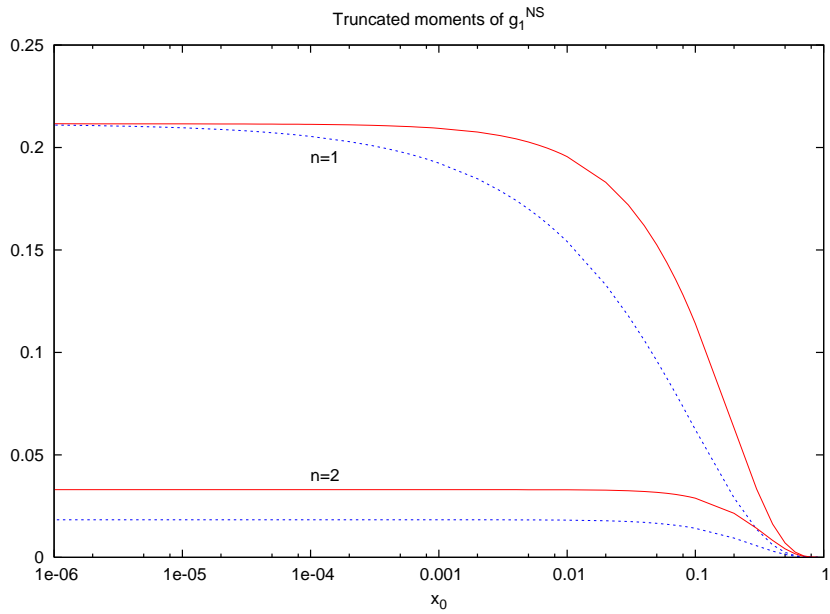}
}
\caption{Truncated $n$-th moments of the spin structure function $g_1^{NS}$ as
a function of the truncation point $x_0$.
The results are shown for the flat (\ref{eq.2.18}) (red solid) and for the
singular (\ref{eq.2.19}) (blue dotted) parametrisation.
The evolution scale $Q^2=10\; {\rm GeV}^2$}
\label{fig:1}
\end{center}
\end{figure}
\begin{figure}
\begin{center}
\resizebox{0.7\textwidth}{!}{
\includegraphics{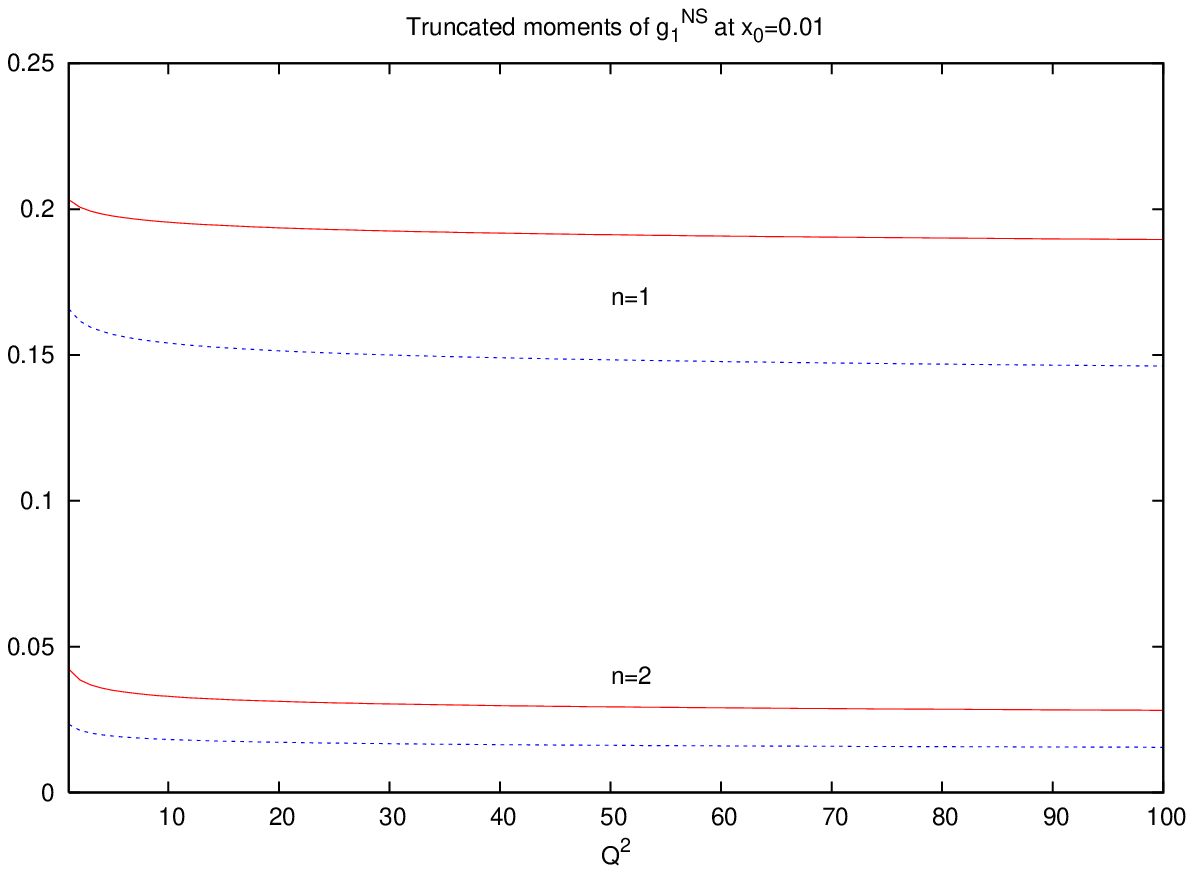}
}
\caption{$Q^2$ evolution of the truncated $n$-th moments of $g_1^{NS}$ at
$x_0=0.01$. The results are shown for the flat (\ref{eq.2.18}) (red solid)
and for the singular (\ref{eq.2.19}) (blue dotted) parametrisation.}
\label{fig:2}
\end{center}
\end{figure}
\begin{figure}
\begin{center}
\resizebox{0.7\textwidth}{!}{
\includegraphics{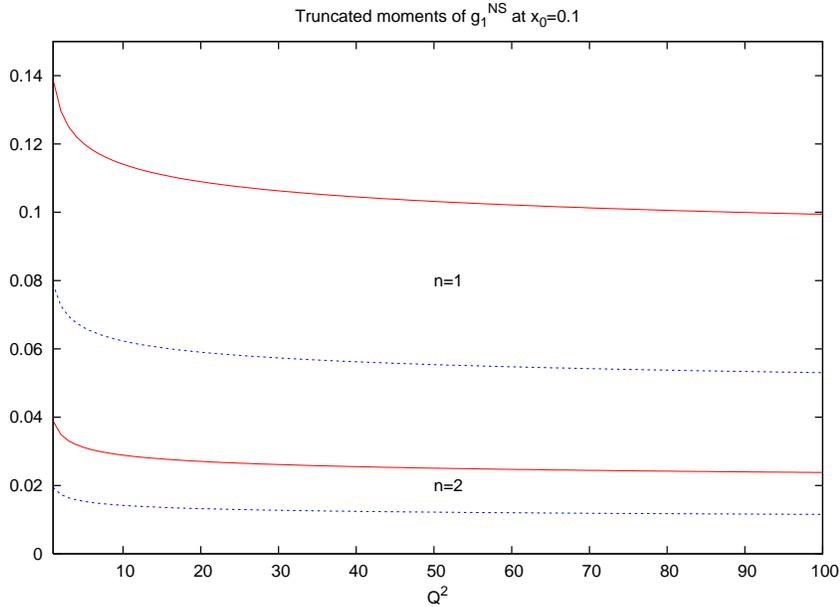}
}
\caption{$Q^2$ evolution of the truncated $n$-th moments of $g_1^{NS}$ at
$x_0=0.1$. The results are shown for the flat (\ref{eq.2.18}) (red solid)
and for the singular (\ref{eq.2.19}) (blue dotted) parametrisation.}
\label{fig:3}
\end{center}
\end{figure}

\section{Summary}

In this paper we have derived DGLAP evolution equations for the truncated
Mellin moments of the parton distributions. We have found that the equations
closely resemble those for the partons themselves. The role of the
splitting function for n-th moment plays $P'(n,x)=x^{n}P(x)$, where $P(x)$
is the well-known splitting function from the DGLAP equation for the partons.
The presented approach has an advantage that it can be successfully used
for each n-th moment and for every truncation point $x_0\in (0;1)$. 
The obtained equations are exact and can be solved with use of standard
methods of solving the DGLAP equations. The polynomial expansion technique
is in the case of the truncated moments more effective than in the case of
the parton densities themselves. This is because $n$-th moments for $n\geq 1$
are usually nonsingular functions of $x$, when $x\rightarrow 0$. In this way
one can obtain reliable results independently of the input parton
distributions.

The truncated moments approach refers directly to the physical values - moments
(rather than to the parton distributions), what enables one to use a wide range
of deep-inelastic scattering data in terms of smaller number of parameters.
In this way, no assumptions on the shape of parton distributions are needed.
Using the evolution equations for the truncated moments one can also avoid
uncertainties from the unmeasurable very small $x\rightarrow 0$ region.

An analysis of the $Q^2$ QCD evolution for the truncated moments of the parton
densities can be a valuable tool e.g. in the determination of the contribution
to different sum rules from the experimentally accessible region. Particularly
important is knowledge on the gluon contribution to the spin of the nucleon.
Thus the truncated moments approach can be useful and interesting both from the
theoretical and the experimental point of view.

\appendix
\section{Detailed derivation of the evolution equations for the truncated
Mellin moments of the parton densities}

Here we show how to obtain eq.(\ref{eq.2.11}). The left-hand side of this
relation can be rewritten in the full form
\begin{equation}\label{eq.A.1}
l.h.s.=\int\limits_{x_0}^1 dx\, x^{n-1}\int\limits_{x}^1 \frac{dz}{z}\,
P\left(\frac{x}{z}\right)\, q(z).
\end{equation}
Using the Heaviside function
\begin{equation}\label{eq.A.2}
\Theta(y)=\cases{1 & for~~ $y>0$ \cr 0 & for~~ $y\leq 0$ \cr},
\end{equation}
we can change the order of integration in (\ref{eq.A.1}), namely
\begin{equation}\label{eq.A.3}
l.h.s.=\int\limits_{x_0}^1 \frac{dz}{z}\, q(z)\int\limits_{x_0}^1 dx\,
x^{n-1}\, \Theta (z-x)\, P\left(\frac{x}{z}\right).
\end{equation}
Now we introduce instead of $x$ a new variable $u$, defined by
\begin{equation}\label{eq.A.4}
u=\frac{x_0 z}{x},
\end{equation}
what implies
\begin{equation}\label{eq.A.5}
l.h.s.=\int\limits_{x_0}^1 dz\, z^{n-1}\, q(z)\int\limits_{x_0}^{z}
\frac{du}{u}\, \left(\frac{x_0}{u}\right)^n \, P\left(\frac{x_0}{u}\right).
\end{equation}
As before, we use the $\Theta$ function and get
\begin{equation}\label{eq.A.6}
l.h.s.=\int\limits_{x_0}^1 dz\, z^{n-1}\, q(z)\int\limits_{x_0}^{1}
\frac{du}{u}\, \Theta (z-u)\,
\left(\frac{x_0}{u}\right)^n \, P\left(\frac{x_0}{u}\right).
\end{equation}
This allows us to change again the order of integration and obtain the final
result
\begin{equation}\label{eq.A.7}
l.h.s.=\int\limits_{x_0}^{1}\frac{du}{u}\,
\left(\frac{x_0}{u}\right)^n \, P\left(\frac{x_0}{u}\right)
\int\limits_{u}^1 dz\, z^{n-1}\, q(z)=(P'\otimes \bar{q}_n)(x_0) = r.h.s.,
\end{equation}
where $P'$ is given by (\ref{eq.2.12}).

\end{document}